\documentclass[aps,showpacs,superscriptaddress,preprint]{revtex4}%
\usepackage{amsfonts}
\usepackage{amsmath}
\usepackage{amssymb}
\usepackage{graphicx}%
\begin{document}
\title{Adsorption and dissociation of H$_{2}$O on Zr(0001) with density-functional
theory studies}
\author{Shuang-Xi Wang}
\affiliation{State Key Laboratory for Superlattices and
Microstructures, Institute of Semiconductors, Chinese Academy of
Sciences, P. O. Box 912, Beijing 100083, People's Republic of China}
\affiliation{Department of Physics, Tsinghua University, Beijing
100084, People's Republic of China}
\affiliation{LCP, Institute of
Applied Physics and Computational Mathematics, P.O. Box 8009,
Beijing 100088, People's Republic of China}
\author{Ping Zhang}
\thanks{Corresponding author; zhang\_ping@iapcm.ac.cn}
\affiliation{LCP, Institute of Applied Physics and Computational
Mathematics, P.O. Box 8009, Beijing 100088, People's Republic of
China}
\author{Peng Zhang}
\affiliation{Department of Nuclear Science and Technology, Xi'an Jiaotong University, Xi'an
710049, People's Republic of China}
\author{Jian Zhao}
\affiliation{State Key Laboratory for Geomechanics and deep underground engineering, China
University of Mining and Technology, Beijing 100083, People's Republic of China}
\author{Shu-Shen Li}
\affiliation{State Key Laboratory for Superlattices and
Microstructures, Institute of Semiconductors, Chinese Academy of
Sciences, P. O. Box 912, Beijing 100083, People's Republic of China}


\begin{abstract}
The adsorption and dissociation of isolated H$_{2}$O molecule on
Zr(0001) surface are theoretically investigated for the first time
by using density-functional theory calculations. Two kinds of
adsorption configurations with almost the same adsorption energy are
identified as the locally stable states, i.e., the flat and upright
configurations respectively. It is shown that the flat adsorption
states on the top site are dominated by the 1$b_{1}$-$d$ band
coupling, insensitive to the azimuthal orientation. The diffusion
between adjacent top sites reveals that the water molecule is very
mobile on the surface. For the upright configuration, we find that
besides the contribution of the molecular orbitals 1$b_{1}$ and
3$a_{1}$, the surface$\rightarrow$water charge transfer occurring
across the Fermi level also plays an important role. The
dissociation of H$_{2}$O is found to be very facile, especially for
the upright configuration, in good accordance with the attainable
experimental results. The present results afford to provide a
guiding line for deeply understanding the water-induced surface
corrosion of zirconium.

\end{abstract}
\maketitle

\section{INTRODUCTION}

The adsorption of water on metal surfaces is of fundamental importance and has
gained a lot of interest associated with a variety of phenomena such as
heterogeneous catalysis and corrosion of materials
\cite{Thiel1987,Henderson2002}. As a result these systems have been
intensively investigated by various experimental and theoretical techniques,
especially for the transition metal surfaces, such as Cu(100)
\cite{Brosseau1993,Sanwu2004}, Fe(100) \cite{Hung1991,Jung2010}, and Pd(100)
\cite{Lloyd1986,Jibiao2007}. From a practical point of view, it is critical to
understand the bonding and orientation characteristics of water molecule on
the surfaces. Experimentally, complicated by the facile H$_{2}$O cluster
formation, it is difficult to discriminate between H$_{2}$O monomers and
clusters. Thus ambiguities have arisen about the preferred orientation of
H$_{2}$O molecule on the surfaces.

Theoretically an upright configuration has been proposed for the adsorption of
water on metal surfaces \cite{Seong1996,Morgenstern2002}. It has been
demonstrated that by maximizing the adsorbate-dipole substrate-image-dipole
interactions, an upright H$_{2}$O favors interaction with the metal surfaces
through the molecular orbitals (MOs) of water, mainly 3$a_{1}$ orbital.
Nevertheless, in later density-functional theory (DFT) calculations, a
flat-lying configuration on the top site of transition metal surfaces has been
established by some sophisticated studies
\cite{Michaelides2003,Sheng2004,Carrasco2009}, arguing that the 1$b_{1}$
orbital dominates the water-surface interaction, by coupling with atomic $d$
orbital of the transition metal surfaces. Much desirable it is to handle these
conflicting results. To this end, further systematic studies in this area are
obviously highly needed for a thorough understanding of the water structures
on metal surfaces.

Motivated by this observation, in the present paper we use first-principles
calculations to investigate the adsorption properties of H$_{2}$O on the
Zr(0001) surface. The reason why we choose zirconium as the prototype is that
zirconium and its alloys have long been used in nuclear reactors as low
neutron adsorption cross-section and excellent corrosion resistance
\cite{stojilovic2005}. Water is the main residual gas in the ultrahigh vacuum
(UHV) vessels of the nuclear reactors, so it is highly meaningful to study the
adsorption of water molecule at zirconium surfaces. The experimental
investigations for the adsorption of water on Zr(0001) have been done with
various techniques, including low-energy electron diffraction (LEED)
\cite{Li1997,Bing1997} and photoemission spectroscopy \cite{Dudr2006}. Water
was found to adsorb on the surface and autocatalytic decomposition took place,
as a function of temperature ($T$) (170 K $<$ $T$ $<$ 573 K). Despite the
experimental results a detailed investigation on the electronic nature of
water adsorption and dissociation on the Zr(0001) surface is indispensable to
a complete understanding of water-metal interactions. Moreover, we expect that
the present work helps to resolve current existing controversy mentioned above.

Through analysis of the electronic projected density of states (PDOS) and
charge density difference, we obtain the adsorption properties of H$_{2}$O on
the Zr(0001) surface. We find that there exist two kinds of adsorption
structures with almost the same adsorption energy as the locally stable
states, including flat and upright configurations respectively. It is found
that dominated by the 1$b_{1}$-$d$ band coupling, the flat adsorption states
on the top site are insensitive to the azimuthal orientation. The diffusion
between adjacent top sites reveals high mobility of the water molecule on
Zr(0001). For the upright configuration, we find that besides the
hybridization contribution of the molecular orbitals 1$b_{1}$ and 3$a_{1}$,
charge transfer between the adsorbate and the substrate near the Fermi energy
also plays an important role in electrostatically stabilizing the adsorption
structure. Consistent with the attainable experimental measurements, the
dissociation of H$_{2}$O is found to be very facile, especially for the
upright configuration.

The rest of the paper is organized as follows. In Sec. II the computational
methods and the supercell models that we use are briefly described. In Sec.
III we present and discuss our results for H$_{2}$O adsorption on the Zr(0001)
surface, followed by diffusion and dissociation properties of the system.
Finally, in Sec. IV, we close our paper with a conclusion of our main results.

\section{COMPUTATIONAL METHODS}

Our calculations are performed within DFT using the Vienna \textit{ab-initio}
simulation package (VASP) \cite{Kresse1996}. The PBE \cite{Perdew1996}
generalized gradient approximation and the projector-augmented wave potential
\cite{Kresse1999} are employed to describe the exchange-correlation energy and
the electron-ion interaction, respectively. The cutoff energy for the plane
wave expansion is set to 400 eV. The Zr(0001) surface is modeled by a slab
composing of five atomic layers and a vacuum region of 20 \AA . A 2 $\times$ 2
supercell, in which each monolayer contains four Zr atoms, is adopted in the
study of the H$_{2}$O adsorption. The water is placed on one side of the slab
only and a dipole correction \cite{Bengtsson1999} is applied to compensate for
the induced dipole moment. During our calculations, the bottom two atomic
layers of the Zr substrate are fixed, and other Zr atoms as well as the
H$_{2}$O molecule are free to relax until the forces on the ions are less than
0.02 eV/\AA . Integration over the Brillouin zone is done using the
Monkhorst-Pack scheme \cite{Monkhorst1976} with 7 $\times$ 7 $\times$ 1 grid
points. And a Fermi broadening \cite{Weinert1992} of 0.1 eV is chosen to smear
the occupation of the bands around the Fermi level by a finite-$T$ Fermi
function and extrapolating to $T$ = 0 K.

The calculation of the energy barriers for the water diffusion and
dissociation processes is performed using the nudged elastic band (NEB) method
\cite{Jonsson1998}, which is a method for calculating the minimum energy path
between two known minimum energy sites, by introducing a number of
\textquotedblleft images\textquotedblright\ along the diffusion path. The
energy barrier is determined by relaxing the atomic positions of each image in
the direction perpendicular to the path connecting the images, until a force
convergence is achieved. In present work, the diffusion path is modeled using
seven images, two of which include the minimum energy sites as initial and
final positions. As an initial guess, five linearly interpolated, intermediate
images between the initial and final configuration are used.

\section{RESULTS AND DISCUSSIONS}

\subsection{ADSORPTION PROPERTIES}

The structural and energetic parameters of the free water molecule are
calculated within a box with the same size of the adsorbed systems. The
optimized geometry for free H$_{2}$O gives a bond length of 0.97 \AA ~ and a
bond angle of 104.2$^{\circ}$, consistent with the experimental values of 0.96
\AA ~ and 104.4$^{\circ}$ \cite{Eisenberg1969}. The calculated lattice
constant of bulk Zr ($a$, $c$) are 3.24 \AA ~and 5.18 \AA , respectively, in
good agreement with the experimental measurements of 3.233 \AA ~and 5.146
\AA \ \cite{Zhao2005}.

\begin{figure}[ptb]
\begin{center}
\includegraphics[width=0.5\linewidth]{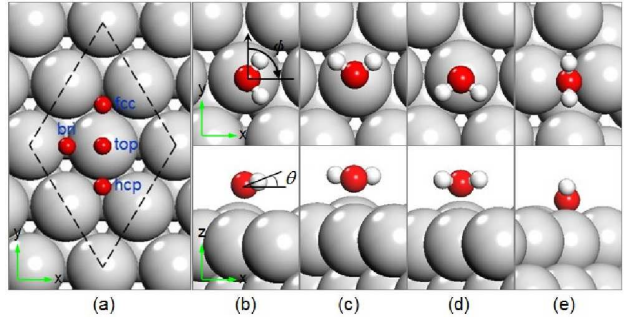}
\end{center}
\caption{(Color online) (a) The structure of the p($2\times2$) surface cell of
Zr (0001), and four on-surface adsorption sites, meanwhile the red balls
denote the initial positions of O atoms in the adsorption picture. (b)-(e) Top
view (upper panels) and side view (lower panels) of the optimized structures
of four most stable adsorption states of H$_{2}$O/Zr(0001) surface, i.e.,
top-$x$, top-$y1$, top-$y2$, and bri-$z$, respectively. Gray, red and white
balls denote Zr, O, and H atoms, respectively.}%
\label{fig1}%
\end{figure}

As depicted in Fig. \ref{fig1}(a), we consider four high-symmetry sites on the
Zr(0001) surface, respectively the top, bridge (bri), hcp and fcc hollow
sites. The O atom of water is initially placed on the precise high-symmetry
sites with various orientations of water with respect to the substrate. We
find that there exist locally stable adsorption states on the top site of
Zr(0001), where the H$_{2}$O molecules lie fairly flat on the surface, labeled
by employing the notations top-$x$, $y1$ and $y2$, respectively. Besides, our
calculations demonstrate that the upright molecular configuration adsorbed on
the bridge site with the O atom lying down is also a locally state molecular
state, labeled by bri-$z$. The structural and energetic details of the
molecular states are illustrated in Figs. \ref{fig1}(b)-(e) and summarized in
Table \ref{table1}. In Fig. \ref{fig1}(b) we define two angles $\phi$ and
$\theta$. $\phi$ represents the azimuthal angle of H$_{2}$O with respect to
the surface, and $\theta$ represents the tilt angle between the H$_{2}$O
molecular dipole plane and the surface. The adsorption energy of the system is
calculated as follows:
\begin{equation}
E_{\mathrm{ad}}=E_{\mathrm{H_{2}O/Zr(0001)}}-E_{\mathrm{H_{2}O}}%
-E_{\mathrm{Zr(0001)}}, \label{Ead}%
\end{equation}
where $E_{\mathrm{H_{2}O}}$, $E_{\mathrm{Zr(0001)}}$, and $E_{\mathrm{H_{2}%
O/Zr(0001)}}$ are the total energies of the H$_{2}$O molecule, the clean Zr
surface, and the adsorption system respectively. According to this definition,
a negative value of $E_{\mathrm{ad}}$ indicates that the adsorption is
exothermic (stable) with respect to a free H$_{2}$O molecule and a positive
value indicates endothermic (unstable) reaction.

\begin{table}[ptbh]
\caption{Calculated structural parameters, adsorption energy for a water
molecule on Zr(0001) surface. E$_{a}$ (eV) represents the adsorption energy.
$\Phi$ (eV) represents the work function. $z_{\text{O}}$ (\AA ) represents the
vertical height of the O atom from the surface, the height of which is
averaged over all atoms on the surface. $d_{\text{O-H}}$ (\AA ) represents the
bond length between the O and H atoms. $\theta$ ($^{\circ}$) represents the
tilt angle between the H$_{2}$O molecular dipole plane and the surface.
$\alpha_{\text{H-O-H}}$ ($^{\circ}$) represents the H-O-H bond angle.}%
\label{table1}
\begin{tabular}
[c]{ccccccc}\hline\hline
\ Site & E$_{a}$ & $\Phi$ & $z_{\text{O}}$ & $d_{\text{O-H}}$ & $\theta$ &
$\alpha_{\text{H-O-H}}$\\\hline
\ top-$x$ & -0.611 & 3.35 & 2.35 & 0.99 & 16.8 & 105.9\\
\ top-$y1$ & -0.616 & 3.38 & 2.35 & 0.99 & 14.9 & 106.2\\
\ top-$y2$ & -0.609 & 3.35 & 2.34 & 0.99 & 15.9 & 106.1\\
\ bri-$z$ & -0.610 & 2.94 & 1.88 & 0.99 & 90.0 & 110.6\\\hline\hline
\  &  &  &  &  &  &
\end{tabular}
\end{table}

From Table \ref{table1}, we can clearly see that at these stable adsorption
sites, the work functions are much smaller than the clean Zr(0001) surface
(4.26 eV), implying an observable charge redistribution between the adsorbate
water and the surface Zr atoms. Taking the adsorption site top-$y1$ for
example, the O-H bond length 0.99 \AA ~is almost identical to 0.97 \AA ~of
free H$_{2}$O molecule, but the H-O-H bond angle 106.2$^{\circ}$ is larger
than that of free H$_{2}$O. The tilt angle is 14.9$^{\circ}$, differing a
little from that ($\sim$10$^{\circ}$) of other transition metal surfaces
mentioned above, where the top site is the most stable state. The adsorption
energy $-$0.616 eV indicates a stronger molecule-surface interaction than on
other transition metal surfaces (usually of $-$0.3 eV). For the adsorption
site bri-$z$, a lower work function 2.94 eV and a larger H-O-H bond angle
110.6$^{\circ}$ are identified, suggesting more prominent charge
redistribution and molecular distortion compared to the top-site adsorption.
It is clear that the bri-$z$ adsorption almost has the same adsorption energy
as the top-site adsorption. This is quite different from previous DFT
calculations on other transition metal surfaces that predicted the flat-lying
top-site adsorption to be the most stable configuration. Interestingly, for a
H$_{2}$O molecule to reach the adsorption state, there does not exist any
energy barrier, which means that H$_{2}$O can be adsorbed on the Zr(0001)
surface spontaneously.

\begin{figure}[ptb]
\begin{center}
\includegraphics[width=0.5\linewidth]{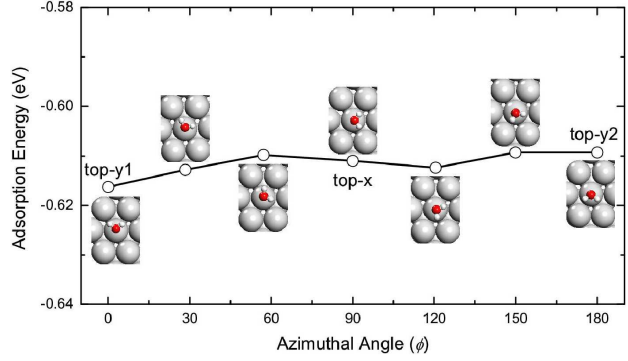}
\end{center}
\caption{(Color online) Calculated adsorption energy of H$_{2}$O as a function
of the azimuthal angle at the top site. The insets show the structures adopted
in the calculations.}%
\label{fig2}%
\end{figure}

As presented in Table \ref{table1}, the energetic differences among these
three adsorption states on top site are very tiny. For further illustration,
we investigate the azimuthal orientation of the adsorbed H$_{2}$O, which is
shown in Fig. \ref{fig2}. It can be seen that the adsorption site top-$y1$ is
slightly more stable. Nevertheless, it is noted that the variation of the
adsorption energy for different azimuthal orientations are determined to be
less than 0.01 eV. We will see below that this is much smaller than the energy
barrier of the lateral diffusion, implying that the azimuthal rotation is
essentially unhindered and may occur at very low temperature.

\begin{figure}[ptb]
\begin{center}
\includegraphics[width=0.5\linewidth]{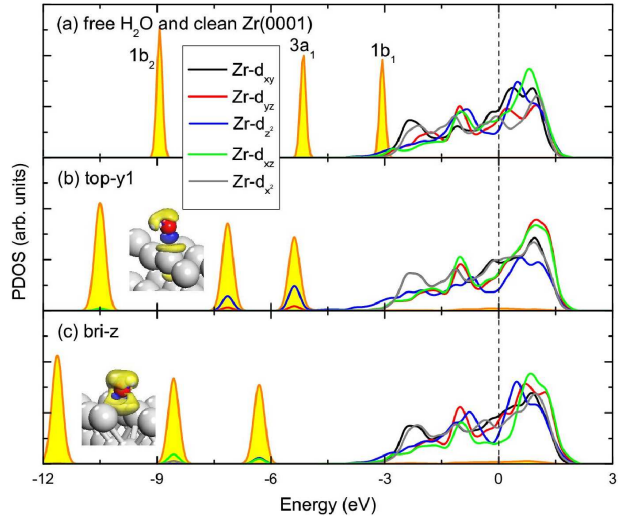}
\end{center}
\caption{(Color online) The PDOS of the H$_{2}$O molecule and the top-layer Zr
atom bonded to H$_{2}$O for (a) free H$_{2}$O and the clean Zr(0001) surface,
(b) top-$y1$ adsorption site and (c) bri-$z$ adsorption site. The insets in
(b) and (c) show the 3D electron density difference, with the isosurface value
set at $\pm$0.015 \textit{e}/\AA $^{3}$. The area filled with yellow color
represents molecular orbital of H$_{2}$O. The Fermi level is set to zero.}%
\label{fig3}%
\end{figure}

In order to further understand the precise nature of the chemisorbed molecular
state, the electronic PDOS of the H$_{2}$O molecule and the topmost Zr layer
are calculated. As typical examples, here we plot in Fig. \ref{fig3} the PDOS
for the stable adsorption configurations of top-$y1$ and bri-$z$. For
comparison, the PDOS of the free H$_{2}$O molecule and clean Zr(0001) surface
are also shown in Fig. \ref{fig3}(a). The three-dimensional (3D) electron
density difference $\Delta\rho(\mathbf{r})$, which is obtained by subtracting
the electron densities of noninteracting component systems, $\rho
^{\text{Be(0001)}}(\mathbf{r})+\rho^{\text{H$_{2}$O}}(\mathbf{r})$, from the
density $\rho(\mathbf{r})$ of the H$_{2}$O/Zr(0001) surface, while retaining
the atomic positions of the component systems at the same location as in
H$_{2}$O/Zr(0001), is also shown in the insets of Fig. \ref{fig3}. Positive
(blue) $\Delta\rho(\mathbf{r})$ indicates accumulation of electron density
upon binding, while a negative (yellow) one corresponds to electron density
depletion. MOs 2$a_{1}$ and 1$b_{2}$ of water (not shown here) are far below
the Fermi level and thus remain intact in water-metal interaction. Here we
consider only three MOs 1$b_{2}$, 3$a_{1}$, and 1$b_{1}$.

In the case of adsorption on the top-$y1$ site, as illustrated in
Fig. \ref{fig3}(b), these three MOs are rigidly shifted downward by
1.57, 2.02 and 2.33 eV, respectively. This is essentially caused by
the different electronegativities of Zr and water molecule, whilch
induces charge redistribution and thus build a global electrostatic
attraction between the water and substrate. In addition to this
rigid energy shift, it is also noticeable that due to the
molecule-metal orbital hybridization, the MOs of adsorbed water are
broadened apparently for 1$b_{1}$ and 3$a_{1}$, which are known to
have an oxygen lone-pair character perpendicular to the molecular
plane, and a mixture of partial lone-pair character parallel to the
molecular plane and partial O-H bonding character, respectively
\cite{Thiel1987,Jung2010}. Remarkably, the water adsorption
introduces new peaks for both $d_{z^{2}}$ and $d_{yz}$ states of the
surface Zr atom, aligning in energy with 1$b_{1}$ and 3$a_{1}$.
Especially for 1$b_{1}$, more electronic states of the Zr atom,
mostly $d_{z^{2}}$ appear nearby, indicating that the adsorbed
1$b_{1}$ MO may act as an electron donor state. This is quite in
accordance with the general picture that the water-surface
interaction is dominated by the 1$b_{1}$-$d$ band coupling. It is
obvious that this kind of coupling cannot be effected essentially by
the azimuthal rotation of the water, hence the adsorption of
H$_{2}$O on the top site is insensitive to the azimuthal
orientation. The features of the orbital hybridization are further
substantiated by the 3D electron density difference plotted in the
inset of Fig. \ref{fig3}(b), from which we can see that there exists
a large charge accumulation between the adsorbate and substrate.

For upright adsorption site bri-$z$ [Fig. \ref{fig3}(c)], it is
known that 3$a_{1}$ plays an additional key role in the upright
adsorption structure of water monomer \cite{Michaelides2003}. Here
obviously, the orbital 3$a_{1}$ undergoes a noticeable broadening as
well as 1$b_{1}$. We notice that instead of $d_{yz}$, it is the
state $d_{xz}$ of the Zr atom, together with $d_{z^{2}}$, that
overlaps with the orbital 1$b_{1}$ of water. And new peaks for
$d_{xz}$ and $d_{x^{2}}$ emerge aligning in energy with the orbital
3$a_{1}$. This observable overlapping can also be seen from the 3D
electron density difference plotted in the inset of Fig.
\ref{fig3}(c), from which we find a large charge accumulation
between the O atom and two adjacent top-layer Zr atoms. Moreover, it
is noteworthy that a discernible occupied domain of states of the
adsorbed water emerges near the Fermi level, suggesting more
prominent charge transfer between adsorbate and substrate than the
top-site adsorption. We find that the emerged occupied state aligns
with the lowest unoccupied MO, which coincides with the so-called
Blyholder model \cite{Blyholder1964}, where an electron donation
from the adsorbate highest occupied MO to substrate states and a
back-donation from such states to the lowest unoccupied MO of the
adsorbate build up the chemisorption bond. With a distortion of the
adsorbed water, the MOs are shifted down by 2.70, 3.43 and 3.25 eV
for 1$b_{2}$, 3$a_{1}$ and 1$b_{1}$, respectively, which are more
pronounced compared with those on top-$y1$, especially for the
orbital 3$a_{1}$, hence more unoccupied MOs are drawn below the
Fermi level. Therefore, although observable is the overlapping,
charge transfer is more prominent for the bri-$z$ adsorption,
leading to a considerable stable electrostatic bonding between water
and the surface.

\subsection{DIFFUSION OF H$_{2}$O MOLECULE}

\begin{figure}[ptb]
\begin{center}
\includegraphics[width=0.5\linewidth]{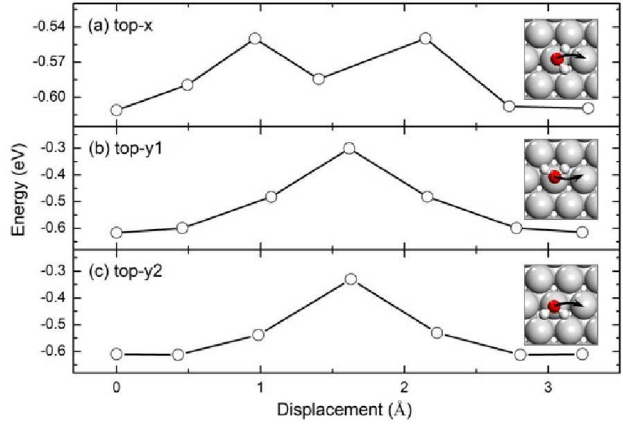}
\end{center}
\caption{(Color online) Diffusion of H$_{2}$O on the top site of Zr(0001)
surface as a function of the lateral displacement of O atom from its original
site top-$x$ (upper panel), top-$y1$ (middle panel), and top-$y2$ (lower
panel), respectively.}%
\label{fig4}%
\end{figure}

Given a H$_{2}$O molecule at a stable adsorption site, it is interesting to
see how it diffuses on the substrate. Here, therefore, we calculate the
diffusion paths and energetic barriers of water on Zr(0001) surface between
neighboring adsorption sites along the top-$x$, top-$y1$ and top-$y2$ channels
respectively, which are schematically shown in the insets of Fig. \ref{fig4}.
Each lateral diffusion path adopted here connects two minimum energy states on
the top sites, through the bridge site as a transition state. The adsorption
energies as a function of the lateral displacement of O atom are shown in Fig.
\ref{fig4}.

For the channel top-$y1$, the transition state with the energy maximum located
on the bridge site, which is not a stable adsorption site. It is found that
the water molecule moves towards the adjacent top-$y1$ site straightforwardly,
with slight wiggling along the diffusion path. Moreover, the diffusion energy
barrier 0.315 eV along this path is much lower than the adsorption energy on
top-$y1$ site, implying that the water molecule has a high mobility on the
Zr(0001) surface. This is the same trend that is found for the channel
top-$y2$, with a lower barrier 0.280 eV.

In the case of channel top-$x$, the lower diffusion barrier 0.061 eV
indicates a higher mobility than that along those fore-mentioned two
channels. Furthermore, there exists a local energy minimum along
this diffusion path. Investigation in more detail in this particular
case reveals that this energy minimum coincides with the stable
bridge adsorption site bri-$z$, which means that the water molecule
may rotate during the diffusion. We find that as the water molecule
migrates towards the adjacent top site, it rotates around the O atom
and overcomes an energy barrier, till the tilt angle reaches as high
as 71$^{\circ}$, where the molecule moves to the bridge site as a
transition state. Then the water molecule rotates reversely,
overcoming another barrier and arriving at the adjacent top-$x$
site. This energy minimum state located on the bridge site may also
give a reason why this channel possesses the lowest diffusion
barrier.

The energy barriers of these lateral diffusion channels are larger than that
of the azimuthal rotations mentioned above, whereas by assuming the attempt
frequency 10$^{13}$ of the adsorbate, the energy barriers of these three
diffusion channels top-$x$, top-$y1$ and top-$y2$ correspond to temperature of
about 24, 122, and 109 K, respectively, suggesting that the diffusion can
occur under room temperature on the H$_{2}$O/Zr(0001) surface, especially for
the channel top-$x$. These results provide evidence that the water molecules
are very mobile on Zr(0001), even at very low temperature \cite{Mitsui2002}.

\subsection{DISSOCIATION OF H$_{2}$O MOLECULE}

Finally let us discuss the possibility of dissociation of H$_{2}$O molecule
into H and OH species on the Zr(0001) surface. In order to investigate the
water dissociation process, we begin with the adsorption properties of the
dissociated H and OH species. The structural and energetic details of these
species are summarized in Table \ref{table2}. For the H species, the hcp site
is the most stable with an adsorption energy of $-$3.298 eV, and the fcc site
is slightly less stable with an adsorption energy of $-$3.240 eV. Next, for
the OH species, the most stable site is found to be fcc with an adsorption
energy of $-$5.627 eV, and the hcp site is less stable by 0.086 eV than the
fcc site. The O-H bond length (0.98 \AA )~with O atom end-on orientation is
less than that (1.00 \AA )~of the free OH molecule.

\begin{table}[ptbh]
\caption{Calculated structural parameters, adsorption energy for water
dissociation products on Zr(0001) surface. E$_{a}$ (eV) represents the
adsorption energy. $z$ (\AA ) represents the vertical height of the H atom
(for H species) or O atom (for OH species) from the surface.}%
\label{table2}
\begin{tabular}
[c]{ccccc}\hline\hline
\ Species & Site & E$_{a}$ & $z$ & $d_{\text{O-H}}$\\\hline
\ H & hcp/fcc & -3.298/-3.240 & 1.13/1.09 & \\
\ OH & fcc/hcp & -5.627/-5.541 & 1.32/1.33 & 0.98/0.98\\\hline\hline
\  &  &  &  &
\end{tabular}
\end{table}

\begin{table}[ptbh]
\caption{Calculated structural parameters, adsorption energy for the final
H+OH configurations and the dissociation barriers of water molecule on
Zr(0001) surface. E$_{d}$ (eV) represents the dissociation barrier.
Site$_{\text{H}}$ and Site$_{\text{OH}}$ represent the positions of H and OH
species in the dissociative configuration, respectively.}%
\label{table3}
\begin{tabular}
[c]{ccccccccc}\hline\hline
\ Path & E$_{a}$ & E$_{d}$ & $\Phi$ & Site$_{\text{H}}$ & Site$_{\text{OH}}$ &
$z_{\text{H}}$ & $z_{\text{O}}$ & $d_{\text{O-H}}$\\\hline
\ top-$x$ & -8.288 & 0.106 & 3.16 & fcc & hcp & 1.07 & 1.34 & 0.97\\
\ top-$y1$ & -8.288 & 0.261 & 3.16 & fcc & hcp & 1.07 & 1.34 & 0.97\\
\ top-$y2$ & -8.462 & 0.371 & 3.18 & hcp & fcc & 1.09 & 1.38 & 0.97\\
\ bri-$z$ & -8.288 & 0.093 & 3.16 & fcc & hcp & 1.07 & 1.34 &
0.97\\\hline\hline
\  &  &  &  &  &  &  &  &
\end{tabular}
\end{table}

We examine four probable dissociation paths of H$_{2}$O molecule on the
Zr(0001) surface. The initial states as the precursors for dissociation are
four stable molecular adsorption states discussed above. For the final states,
based on the results of the adsorption of H and OH, we consider several
possible combinations of the adsorption sites for H and OH species. It is
found that for the H+OH configuration, the H and OH species occupy two
neighboring hollow sites (fcc and hcp) respectively connected via a bridge
site, which is the most stable. Table \ref{table3} shows the structural and
energetic details of the final H+OH configurations. We can see from the low
adsorption energies and work functions that these two species strongly bond to
the surface, with OH oriented almost perpendicular to the surface. The O$-$H
bond length 0.97 \AA ~differs only slightly from that of the adsorbed H$_{2}$O molecule.

\begin{figure}[ptb]
\begin{center}
\includegraphics[width=0.5\linewidth]{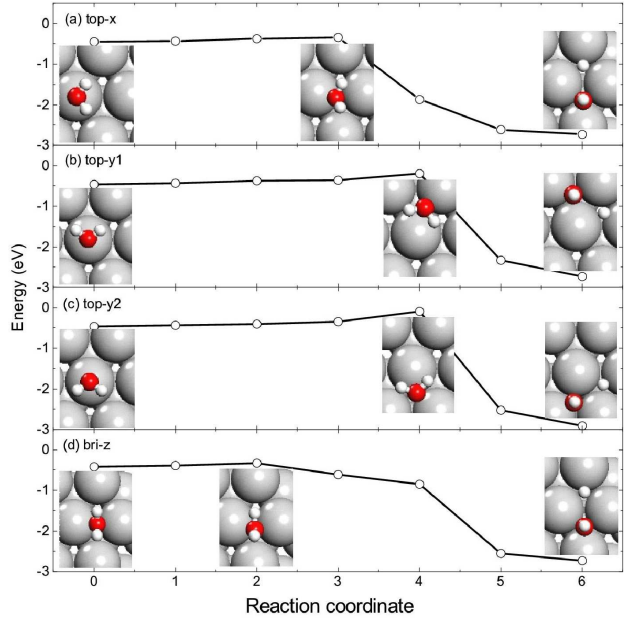}
\end{center}
\caption{(Color online) Four water dissociation paths considered in this
study, corresponding to the four most stable adsorption states represented in
Fig. 1. The inset pictures show the structures of H$_{2}$O/Zr(0001) surface
corresponding to the adsorption, transition and dissociation states,
respectively.}%
\label{fig5}%
\end{figure}

Figure \ref{fig5} shows the energy profiles for these dissociation paths. The
activation energies for the dissociation of the water molecules are 0.093,
0.106, 0.261 and 0.371 eV along the paths bri-$z$, top-$x$, top-$y1$ and
top-$y2$, respectively (see Table \ref{table3}). We find that the paths
top-$x$ and bri-$z$ have noticeably low dissociation barriers, small enough to
facilitate the dissociations of the adsorbed water molecules. While for the
paths top-$y1$ and top-$y2$, the energy barriers are higher than that of the
forenamed two paths. The reason why such differences exist about the energy
barriers will be clear when we go deep into the dissociation process. For the
path bri-$z$, it can be seen (from the inset pictures of Fig. \ref{fig5}) that
the water molecule starts to dissociate almost \textit{in situ}, hence the
energy barrier is low. In the case of path top-$x$, the water molecule first
migrates to the adjacent bridge site, meanwhile the molecule rotates around
the O atom for a larger tilt angle (which it can do with little energy loss),
then following the dissociation process just as in the path bri-$z$. On the
other hand, for the dissociation paths top-$y1$ and top-$y2$, before reaching
the transition states, the water molecules have to rotate more complexly, then
enter the dissociation process with the H and OH species moving to their final
positions. By assuming the attempt frequency 10$^{13}$ of the adsorbate, the
energy barriers of these four dissociation paths correspond to temperature of
about 36, 41, 101 and 144 K, respectively, suggesting that the dissociation
can occur under room temperature on the H$_{2}$O/Zr(0001) surface, especially
for the path bri-$z$. This is in good accordance with the experimental
observations that water can spontaneously dissociate on the Zr(0001) surface
\cite{Bing1997,Dudr2006}, but different from the dissociation of water on
other transition metal surfaces such as Fe(100) \cite{Jung2010}, Rh(111) and
Ni(111) \cite{Pozzo2007}, which need to overcome much higher barriers.

\section{CONCLUSIONS}

In conclusion, we have systematically studied the adsorption and
dissociation behaviors of H$_{2}$O on the Zr(0001) surface by using
first-principles DFT method. Two kinds of adsorption structures with
almost the same adsorption energy were identified as the locally
stable states, i.e., the flat and upright configurations
respectively. It has been shown that the flat adsorption states on
the top site are dominated by the interaction between the water MO
1$b_{1}$ and the metal $d$ band, insensitive to the azimuthal
orientation. The diffusion of water across the surface reveals that
the adsorbed water is very mobile on the surface. For the upright
configuration, we have found that the hybridization between the $d$
band of the Zr(0001) surface and MOs 1$b_{1}$ and 3$a_{1}$ of the
water, as well as the charge transfer between the adsorbate and the
substrate, contribute to the adsorption system. Consistent with
previous experimental results, the dissociation of H$_{2}$O has been
found to be very facile, especially for the upright configuration.
We expect that the present results are greatly helpful for the
practical usage of zirconium in nuclear reactors.

\end{document}